\documentclass[11pt]{article}

\usepackage{amsmath}
\usepackage{amsfonts}
\usepackage{amssymb}
\usepackage{graphicx}
\usepackage{subfigure}
\usepackage{rotating}
\usepackage{color}
\usepackage{amsthm}
\usepackage{authblk}
\usepackage{threeparttable}
\usepackage{pdflscape}
\usepackage{multirow}
\usepackage{bm}
\usepackage{natbib}
\usepackage[ruled]{algorithm2e}

\allowdisplaybreaks[3]
\newcounter{remarkCounter}
\newtheorem{remark}[remarkCounter]{Remark}

\theoremstyle{plain}
\newtheorem{theorem}{Theorem}

\usepackage{setspace}

\begin{document}
\begin{spacing}{1.0}

\title{A Robust Approach for Identifying Gene-Environment Interactions for Prognosis}
\author{Hao Chai$^{1}$, Qingzhao Zhang$^{2}$, Yu Jiang$^{1}$, Guohua Wang$^{2}$, Sanguo Zhang$^{2}$, Shuangge Ma$^{1}$\\
$^{1}$Department of Biostatistics, Yale University;
$^2$School of Mathematical Sciences, University of Chinese Academy of Sciences
}

\maketitle

\noindent
{\bf Running title}: A Robust Approach for Identifying G-E interactions\\
{\bf Contact}: Shuangge Ma\\
60 College ST, New Haven, CT 06520\\
Email: shuangge.ma@yale.edu;
Tel: 1-203-785-3119; Fax: 1-203-785-6912\\

\begin{abstract}
For many complex diseases, prognosis is of essential importance. It has been shown that, beyond the main effects of genetic (G) and environmental (E) risk factors, the gene-environment (G$\times$E) interactions also play a critical role. In practice, the prognosis outcome data can be contaminated, and most of the existing methods are not robust to data contamination. In the literature, it has been shown that even a single contaminated observation can lead to severely biased model estimation. In this study, we describe prognosis using an accelerated failure time (AFT) model. An exponential squared loss is proposed to accommodate possible data contamination. A penalization approach is adopted for regularized estimation and marker selection. The proposed method is realized using an effective coordinate descent (CD) and minorization maximization (MM) algorithm. Simulation shows that without contamination, the proposed method has performance comparable to or better than the unrobust alternative. With contamination, it outperforms the unrobust alternative and, under certain scenarios, can be superior to the robust method based on quantile regression. The proposed method is applied to the analysis of TCGA (The Cancer Genome Atlas) lung cancer data. It identifies interactions different from those using the alternatives. The identified marker have important implications and satisfactory stability.
\end{abstract}


\clearpage
\section*{Introduction}

For many complex diseases such as cancer, diabetes, and cardiovascular diseases, prognosis is of essential interest. In the omics era, profiling studies have been extensively conducted, searching for genetic markers associated with prognosis. It has been suggested that, beyond the main effects of genetic (G) and environmental (E) risk factors, the gene-environment (G$\times$E) interactions also have important implications. Multiple statistical methods have been developed for G$\times$E interaction analysis. For reviews, refer to~\cite{Caspi2006gene},~\cite{Cordell2009detecting},~\cite{Thomas2010methods}, and others.

Denote $T$ as the prognosis time of interest, $X = (X_1, \ldots, X_q)$ as the $q$ environmental/clinical variables, and $Z=(Z_1, \ldots, Z_p)$ as the $p$ genetic variables. Assume $n$ independent subjects. Regression-based interaction analysis, with its broad applicability, has been extensively adopted and proceeds as follows. (a) For gene $j (=1, \ldots, p)$, consider the model $T \sim \phi( \alpha_{j,0}+\sum_{k=1}^qX_k\alpha_{j,k} + Z_j\beta_j + Z_j\sum_{k = 1}^qX_k\gamma_{j,k})$, where $\phi(\cdot)$ is the known link function (for example, the Cox or exponential model), and $\alpha_{j,0},\alpha_{j,k}, \beta_{j}, \gamma_{j,k}$ are the unknown regression coefficients. As usually $q<<n$, this is a low-dimensional model and can be fitted using standard, usually likelihood-based, techniques and software. Denote $p_{j,k}$ as the p-value for $\gamma_{j,k}$. (b) With $\{p_{j,k}: j=1, \ldots, p; k=1, \ldots, q \}$, conduct multiple comparison adjustment using the Bonferroini or FDR (false discovery rate) approach, and identify important interactions.

A limitation of the above approach is a lack of robustness properties. Usually it is assumed that all subjects satisfy the same prognosis models. In practice, most genetic studies cannot afford to conduct strict subject selection. Seemingly homogeneous subjects can have different disease subtypes~\citep{Haibe-Kains2012three} and different survival patterns~\citep{burgess2011cancer}. Cause of death can be misclassified, leading to contamination in disease-specific survival~\citep{fall2008reliability}. The survival times extracted from medical records are not always reliable (\cite{bowman2011doctors},~\cite{rampatige2013assessing}). With unrobust for example likelihood-based estimation, even a single contaminated observation can result in severely biased estimates~\citep{huber2009robust}, which can lead to false marker identification. Another limitation is that, significance level-based identification, although {\it asymptotically} valid, may generate unreliable results when the sample sizes are small to moderate, as in typical profiling studies. A recent study suggests that regularized estimation can lead to more reliable estimation and hence more accurate marker identification~\citep{shi2014gene}.

With low-dimensional biomedical data, robust statistical methods have demonstrated great power. As suggested in a recent review by~\cite{wu2014selective}, with genetic data, the development is limited and unsystematic and has been mostly on the analysis of main effects but not interactions. In the literature, relevant studies include~\cite{gui2011novel}, which identifies important interactions using the multifactor dimensionality reduction (MDR) technique. However, this method is limited to categorial data (such as SNPs) and not broadly applicable.~\cite{shi2014gene} developed a rank-based method, which is robust to model mis-specification but not data contamination. The most relevant study is perhaps~\cite{wang2015identifying}, which developed a quantile-regression based method. With that method, the quantile needs to be specified, which is not a trivial task in practical data analysis. The objective function is not differentiable, causing difficulty in estimation. In addition, as to be shown in this study, its numerical performance can be less satisfactory under many data settings. With low-dimensional data, it has been shown that no robust method dominates the others. It is thus prudent to develop alternative robust methods.

Consider prognosis data with both G and E measurements. The goal is to develop a new method for identifying important G$\times$E interactions. Significantly advancing from the existing studies, the proposed method is robust to contamination in the prognosis data. In addition, our simulation suggests that, under certain scenarios, its numerical performance is better than the quantile regression analysis, which is perhaps the most popular robust method for genetic data~\citep{wu2014selective}. A penalization approach is adopted for marker identification, which differs from the significance level-based approach and can have better numerical performance when the sample size is small to moderate.

\section*{Robust identification of interactions}

\subsection*{Data and model settings}

In the literature, there are multiple prognosis models. In this study, we consider the accelerated failure time (AFT) model, which has been adopted in~\cite{shi2014gene}, \cite{liu2013identification}, and many others. Compared to alternatives such as the Cox model, the AFT model can be preferred because of its more intuitive interpretations and lower computational cost, which are especially desirable with high-dimensional genetic data. With a slight abuse of notation, still $T$ to denote the logarithm of prognosis time. For gene $j$, the AFT model assumes that
$$T = \alpha_{j,0}+\sum_{k=1}^qX_k\alpha_{j,k} + Z_j\beta_j + Z_j\sum_{k = 1}^qX_k\gamma_{j,k}+\epsilon,$$
where $\epsilon$ is the random error.

Assume $n$ independent observations. Consider the scenario where a small subset of the random errors are contaminated, leading to contamination in the prognosis times. In a typical profiling study, $q<<n$, while $p$ can be comparable to or much larger than $n$. For subject $i$, denote $C_i$ as the logarithm of censoring time and $\mathbf{x}_i$ and  $\mathbf{z}_i$ as the observed $X$ and $Z$ values, respectively. Under right censoring, we observe $(y_i=min(T_i, C_i), \delta_i=I(T_i \leq C_i), \mathbf{x}_i, \mathbf{z}_i)$. Further denote $\mathbf{u}_{i,j}=(1,\mathbf{x}^{\prime}_i, z_{i,j},z_{i,j}x_{i,1},\ldots,z_{i,j}x_{i,q})'$,  $\bm\zeta_j=(\alpha_{j,0},\alpha_{j,1},\ldots,\alpha_{j,q},\beta_j,\gamma_{j,1},\ldots, \gamma_{j,q})'$, and $\mathbf{U}_j=(\mathbf{u}_{1,j}, \ldots, \mathbf{u}_{n,j})'$ which is a $n\times (2q+2)$ matrix. For gene $j$ and subject $i$, the AFT model can now be written as
$T_i = \mathbf{u}_{i,j}^\top\bm{\zeta}_j + \epsilon_{i,j}.$ Without loss of generality, assume that the data $\{(y_i,\delta_i,\mathbf{x}_i,\mathbf{z}_i), i=1,\ldots, n\}$ have been sorted according to $y_i$'s from the smallest to the largest.

\subsection*{Penalized robust identification}

A penalized marker identification method is defined by its loss function and penalty, which are defined separately as follows.

\noindent{\bf Loss function} Before defining the robust loss function, we take one step back and consider the scenario with no data contamination. When the distribution of $\epsilon$ is not specified, likelihood-based estimation cannot be adopted. A popular approach is the weighted least squared estimation proposed in~\cite{Stute1993consistent} and proceeds as follows. First compute the Kaplan-Meier weights as $$\omega_{1}=\frac{\delta_{1}}{n},~\omega_{i}=
\frac{\delta_{i}}{n-i+1}\prod_{j=1}^{i-1}
(\frac{n-j}{n-j+1})^{\delta_{j}},~i=2,\cdots,n.$$
For gene $j(=1,\ldots, p)$, the weighted least squared objective function is defined as
\begin{eqnarray}
\sum_{i=1}^n \omega_{i} (y_i - \mathbf{u}_{i,j}^\top\bm\zeta_j)^2,
\end{eqnarray}
This loss function has the following properties. With a simple linear regression model, the most commonly adopted loss function is the least squared one. Here to accommodate censoring, a weight function is imposed to re-weigh different observations according to their observed times and event status. When there is no censoring, $\omega_i=1/n$. With the quadratic form, the loss function is not robust to data contamination. If subject $i$ is not censored, then $\omega_i\neq 0$, and an arbitrarily large $y_i$ results in arbitrarily large estimates. Biased estimation can happen with data contamination, which may lead to false marker identification.

To accommodate data contamination, for gene $j(=1,\ldots, p)$, we propose the exponential squared loss function
\begin{eqnarray}
Q_{\theta}(\bm\zeta_j|\mathbf{y}, \mathbf{U}_j,\bm\omega)=
\sum_{i=1}^n\omega_i\exp(-(y_i - \mathbf{u}_{i,j}^\top\bm\zeta_j)^2 / \theta).
\end{eqnarray}
$\mathbf{y}$ and $\bm\omega$ denote the vectors composed of $y_i$'s and $\omega_i$'s, respectively. $\theta>0$ is a tuning parameter. The rationale of this approach is as follows. For contaminated subjects with $y_i$s deviate from $\mathbf{u}_{i,j}^\top\bm\zeta_j$s (predicted values from the model), $(y_i - \mathbf{u}_{i,j}^\top\bm\zeta_j)^2$s have large values.
The exponential function down-weighs such contaminated observations. The degree of down-weighing is adjusted by $\theta$: with $\theta$ getting smaller, contaminated observations have smaller influence. To accommodate censoring, $\omega_i$'s are imposed in a similar manner as the original Stute's approach. For low-dimensional linear regression model without censoring, the exponential squared loss has been examined in~\cite{wang2013robust}. Different from the existing studies, here we consider the more challenging high-dimensional genetic data especially interactions. In addition, the Kaplan-Meier weights are introduced to accommodate censoring. As to be shown in Appendix, such differences lead to significant differences in the statistical development.

\noindent{\bf Penalized estimation}
For gene $j(=1,\ldots, p)$, we propose the penalized objective function
\begin{eqnarray}
L_{\lambda,\theta}(\bm\zeta_j|\mathbf{y}, \mathbf{U}_j,\bm\omega) = Q_{ \theta}(\bm\zeta_j|\mathbf{y}, \mathbf{U}_j,\bm\omega)-\lambda||\bm\zeta_j||_1.
\label{eqn2}
\end{eqnarray}
$\lambda>0$ is the data-dependent tuning parameter, and $||\cdot||_1$ is the $\ell_1$ norm. Denote $\tilde{\bm\zeta_j}$ as the maximizer of $L_{\lambda,\theta}(\bm\zeta_j|\mathbf{y}, \mathbf{U}_j,\bm\omega)$. Interactions (and main effects) corresponding to the nonzero components of $\tilde{\bm\zeta_j}$ are identified as important.

The strategy of using penalization for identifying important interactions has been adopted in~\cite{shi2014gene} and other studies. Here a Lasso penalty is imposed, which is the most commonly adopted penalty. $\lambda$ controls the degree of sparsity and number of identified interactions. We impose the same $\lambda$ to all genes to ensure comparability.

Multiple penalties can take place of Lasso. In some recent studies such as~\cite{bien2013lasso} and~\cite{liu2013identification}, penalties have been developed to respect the ``main effects, interactions" hierarchy, which reinforces that the main effects corresponding to the identified interactions must be identified. In a large number of studies (\cite{zimmermann2011interaction},~\cite{Caspi2006gene}), it has been observed that genes can have important G$\times$E interactions but no main effects. In addition, if the hierarchy has to be reinforced, we can identify the important interactions first, and then add back the corresponding main effects. Computationally, Lasso is much simpler than the existing alternatives. Our limited experience suggests that with the complex robust loss function, complex penalties have a considerable probability of running into convergence problems. With the above considerations, the Lasso penalty is adopted here.

\noindent{\bf Consistency properties} A significant advantage of the proposed method is that it has the much deserved consistency properties under the ultrahigh dimensional setting. For many of the existing interaction analysis methods, there is a lack of theoretical development. The rigorously established consistency properties provide a solid statistical ground for the proposed method. Details are provided in Appendix.

\subsection*{Computation}

We propose a coordinate-wise updating procedure to compute the solution to~\eqref{eqn2}. For low-dimensional data under simpler settings, an iterative approach is suggested in~\cite{wang2013robust} to select the robust tuning parameter $\theta$. However, under the present high-dimensional settings and with the coordinate-wise updating procedure, such an approach is computationally infeasible. Alternatively, for each $(\lambda,\theta)$ pair, we compute the solution to each marginal model. This way, we can obtain a solution surface over the two-dimensional tuning parameter grid. Then a prediction error based method such as cross-validation can be used to select the robust and penalization tuning parameters. Computation is conducted for each gene separately and can be realized in a highly parallel manner to reduce computer time.

Consider gene $j(=1,\ldots, p)$. Let $r_i(\bm\zeta_j)=y_i-\mathbf{u}_{i,j}^\top\bm\zeta_j$. The first and second order derivatives of $Q(\bm\zeta_j)$ are
\begin{equation}
\begin{cases}
\dot{Q}_k(\bm\zeta_j)=\frac{\partial Q(\bm\zeta_j)}{\partial \zeta_{j,k}}=2\sum_{i=1}^n\omega_iu_{(i,j)_k}r_i(\bm\zeta_j)
\exp(-r_i^2(\bm\zeta_j)/\theta) /\theta, \\
\ddot{{Q}}_{kl}(\bm\zeta_j)=\frac{\partial^2 Q(\bm\zeta_j)}{\partial \zeta_{j,k}\partial \zeta_{j,l}}=2\sum_{i=1}^n\omega_iu_{(i,j)_k}u_{(i,j)_l}\exp(-r_i^2(\bm\zeta_j)/\theta) (2r_i^2(\bm\zeta_j)/\theta-1)/\theta.
\end{cases}
\label{eqn4}
\end{equation}
For a $\bm{\zeta}_j^m$ that is in a small neighborhood of $\bm{\zeta}_j$,  $Q(\bm\zeta_j)$ in~\eqref{eqn2} can be locally approximated by
\begin{equation}
Q(\bm\zeta_j) \approx Q(\bm\zeta^m_j) + \dot{Q}(\bm\zeta^m_j)^\top(\bm\zeta_j-\bm\zeta_j^m)+
\frac{1}{2}(\bm\zeta_j-\bm\zeta^m_j)^\top\ddot{Q}
(\bm\zeta^m_j)(\bm\zeta_j-\bm\zeta^m_j).
\label{eqn5}
\end{equation}
Replacing $Q(\bm\zeta_j)$ in~\eqref{eqn2} with this approximation and taking the first order derivative of $L(\bm\zeta_j)$ with respect to the $k$th element $\zeta_{j,k}$ give us
\begin{equation}
\zeta_{j,k}^{m+1}=\zeta_{j,k}^m -\ddot{Q}^{-1}_{kk}(\bm{\zeta}^m_j)\dot{Q}_k(\bm{\zeta}^m_j)+ \ddot{Q}_{kk}^{-1}(\bm{\zeta}^m_j)\lambda.
\end{equation}
Note that when $2r_i^2(\bm{\zeta}_j)/\theta>1$, $\ddot{Q}_{kk}(\bm{\zeta}_j)\geq0$. Then \eqref{eqn2} can be maximized at infinity, and the algorithm may fail to converge. To avoid this problem, notice that $$\ddot{Q}_{kk}(\bm{\zeta}_j)\geq -2\sum_{i=1}^n\omega_iu_{(i,j)_k}^2\exp(-r_i^2(\bm\zeta_j)/\theta)/\theta$$ for $\omega_i\geq 0$. The right hand side is always non-positive. We re-define $$\ddot{Q}_{kk}(\bm{\zeta}_j)\equiv -2\sum_{i=1}^n\omega_iu_{(i,j)_k}^2\exp(-r_i^2(\bm\zeta_j)/\theta)/\theta,$$ and use the minorization-maximization (MM) algorithm to ensure convergence. The algorithm that combines the coordinate descent and MM algorithm is summarized in Algorithm~\ref{alg1}.
\begin{algorithm}
\caption{The coordinate descent MM algorithm for marginal model $j$}
\label{alg1}
\SetKwInOut{Input}{input}\SetKwInOut{Output}{output}
\Input{$\mathbf{y}, \bm{\omega}, \mathbf{U}_j$, and tuning parameters $\lambda$ and $\theta$.}
\Output{the maximizer $\tilde{\bm{\zeta}}_j$ defined in~\eqref{eqn2}.}
\BlankLine
\textbf{initialization}: let $m = 0, \bm{\zeta}^0_j=\mathbf{0}$. Normalize $\mathbf{y}$ and $\mathbf{U}_j$ such that $\sum_{i=1}^n\omega_iy_i=0$, $\sum_{i=1}^n\omega_i u_{(i,j)_k}=0$ and $\sum_{i=1}^n\omega_iu_{(i,j)_k}^2=n$ for $k=1,\ldots,2q+2$.\;
\Repeat{$||\bm{\zeta}^{m_0}_j - \bm{\zeta}^{m}_j||_2\leq$ a predefined thresholding value}{
$m_0 = m$\;
\For{$k = 1, 2, \ldots, 2q+2$}{
\Repeat{$|{\zeta}^{m}_{j,k} - {\zeta}^{m-1}_{j,k}|\leq$ a predefined thresholding value. In our numerical study, we set the thresholding =$10^{-3}$}{
$\zeta_{j,k}^{m+1} = \zeta_{j,k}^m - \ddot{Q}^{-1}_{kk}(\bm{\zeta}^m_j)\dot{Q}_k(\bm{\zeta}^m_j)+ \ddot{Q}_{kk}^{-1}(\bm{\zeta}^m_j)\lambda$\;
$\zeta_{j,l}^{m+1} = \zeta_{j,l}^m$ for $l = 1, 2, \ldots, 2q+2$, $l\neq k$\;
$m = m + 1$\;
}
}
}
\Return{$\bm{\zeta}^m_j$} as the maximizer.
\end{algorithm}

As mentioned above, we search over a two-dimensional grid of $(\lambda,\theta)$ for the optimal. The range of $\lambda$ is determined as follows. First, its upper bound $\lambda_{\max}$ is selected such that $\tilde{\bm{\zeta}}_j=\mathbf{0}$ for $j=1,\ldots,p$. With the unrobust weighted least squared loss, $\lambda_{\max} = \max_{j=1}^p\{||\mathbf{U}_j^\top W \mathbf{y}||_{\infty}\}$ where $W$ is the diagonal matrix composed of $\omega_i$s. With the proposed robust method, the derivatives in~\eqref{eqn4} can be viewed as a weighted sum of $u_{(i,j)_k}r_i(\bm{\zeta}_j)$s. Because the weight for each subject changes with $\bm{\zeta}_j$, the previously defined $\lambda_{\max}$ may not guarantee that $\tilde{\bm{\zeta}}_j=\mathbf{0}$ for $j$. After some trials, we find that $\lambda_{\max} = 20\max_{j=1}^p\{||\mathbf{U}_j^\top W \mathbf{y}||_{\infty}\}$ is in general a safe upper bound for $\lambda$. The lower bound is chosen as $\lambda_{\min}=\lambda_{\max} / 1000.$ The range of $\theta$ depends on the degree of contamination in data, which is not known in practice. In our numerical studies, we find that the estimator is not very sensitive to the value of $\theta$. For cautions, we choose a relatively wide range for $\theta$. Specifically, after centralization, we choose $\theta \in (\min_{i=1}^n{y_i^2}/100, \max_{i=1}^ny_i^2 \times 100)$.

With the quadratic approximation and minorization, optimization in~\eqref{eqn2} is non-convex. Therefore, there is a risk of converging to local maximums. Thus when trying to solve the estimate at a new $(\lambda, \theta)$ point, instead of using $\bm{\zeta}^0_j=\mathbf{0}$ as the starting value, we use the estimate at a neighboring $(\lambda, \theta)$ point as the warm starting value. This modification speeds up the computation and also improves the convergence property.

\section*{Simulation}

In simulation, we set $n= 300$, $q=3$, and $p = 500, 1000$. There are a total of 18 nonzero effects: 3 main effects of E factors, 5 main effects of G factors, and 10 interactions. The positions of nonzero main G effects and interactions are uniformly placed. The nonzero regression coefficients are randomly generated from $uniform(0.5, 1.5)$. The E and G factors are generated from multivariate normal distributions with marginal means zero, marginal variances one, and the following variance matrix structures: Independent, AR(0.2), AR(0.8), Band(0.3), and Band(0.6). Under the AR($\rho$) correlation structure, for the $i$th and $j$th factors, $corr=\rho^{|i-j|}$. Under the Band($\rho$) correlation structure, for the $i$th and $j$th factors, $corr=\rho I(|i-j|\leq 2)$, where $I(\cdot)$ is the indicator function. Under each correlation scenario, consider seven different distributions for the random error $\epsilon$: N(0,1), 0.95N(0,1)+0.05Cauchy, 0.85N(0,1)+0.15Cauchy, 0.7N(0,1)+0.3Cauchy, 0.95N(0,1)+0.05t(3), 0.85N(0,1)+0.15t(3), and 0.7N(0,1)+0.3t(3). That is, we consider the scenarios with no contamination and three different levels of contamination. Two contamination distributions are considered with different thickness of tails. The log event times are generated from the AFT models. The censoring times are generated independently from exponential distributions. The parameters are adjusted so that the censoring rates are about 25\%.

Beyond the proposed method (referred to as ``Robust"), we also consider the following three alternatives: (a) the ``Unrobust" method, which adopts the weighted least squared loss function and applies the Lasso penalization for selecting important effects; (b) the ``Stute" method, which adopts the weighted least squared loss function, does not apply any penalization, and uses the significance level (p-value) as the criterion for quantifying the importance of effects; and (c) the ``Quantile" method, which adopts the quantile regression-based robust loss function and applies the Lasso penalization for selecting important effects. The Unrobust and Stute methods, as the proposed, are built on the weighted least squared estimation. Comparing with them can establish the advantage of robust loss and penalization-based identification, respectively. The Quantile method is also robust and has been recently developed. Comparing with it can establish the advantage of the weighted exponential squared loss. We acknowledge that multiple methods are potentially applicable to the simulated data. The three alternatives have frameworks closest to that of the proposed.

With the Robust, Unrobust, and Quantile methods, the numbers of selected interactions depend on the tuning parameter values. With the Stute method, the number depends on the p-value cutoff. To eliminate the effect of tuning parameter selection on identification, we examine a sequence of tuning parameter values, evaluate identification performance at each value, and use the ROC (receiver operating characteristics) curve to evaluate the interaction identification accuracy of different methods. A representative ROC plot is shown in Figure 1. In this plot, the proposed method has the dominatingly better accuracy.

Summary AUCs based on 100 replicates are shown in Tables 1 ($p=500$) and 2 ($p=1000$). When there is no contamination, performance of the proposed method is comparable to or slightly worse than that of the unrobust method. For example with $p=500$ and the independence correlation structure, the robust and unrobust methods have mean AUCs 0.861 and 0.889, respectively. And with the AR(0.2) correlation, they have mean AUCs 0.901 and 0.881, respectively. With contamination, the robust method outperforms the unrobust method. For example with $p=500$, the AR(0.2) correlation structure, and 0.7N(0,1)+0.3Cauchy error, the robust and unrobust methods have mean AUCs 0.886 and 0.751, respectively. Under all simulation scenarios, the Stute method, which adopts the robust loss function but significance level-based identification, has inferior identification accuracy. As has been suggested in the literature, with a moderate sample size, the unregularized estimates can be less reliable, leading to inaccurate identification. When comparing the proposed method with the quantile regression-based, we see that under the majority of the settings, the proposed method has superior performance. For example with $p=500$, AR(0.2) correlation, and 0.85N(0,1)+0.15Cauchy error, the proposed method and Quantile have mean AUCs 0.892 and 0.842, respectively. However, under a small number of settings, the Quantile method excels. For example with $p=500$, AR(0.8) correlation, and 0.7N(0,1)+0.3t(3) error, the two methods have mean AUCs 0.863 and 0.933, respectively. Such observations are also reasonable. No robust approach is expected to be able to dominate all others. The proposed method outperforms the quantile regression-based under most scenarios and provides a useful alternative.

\section*{Analysis of the lung squamous cell carcinoma data}

Lung squamous cell carcinoma is the second most common lung cancer and causes around 400,000 deaths each year worldwide~\citep{comprehensive2012TCGA}. Profiling studies have been extensively conducted searching for its prognostic factors. In this section, we analyze the TCGA (The Cancer Genome Atlas) data on the prognosis of lung squamous cell carcinoma. The TCGA data were recently collected and published by NCI and have a high quality. The dataset we analyze was downloaded in April of 2015.

The prognosis outcome of interest is overall survival. Multiple environmental and clinical variables are available for analysis. We first remove variables with a low quality of measurements (for example with a high missing rate). We then select the following variables for analysis: age, gender (female is coded as baseline), smoking pack years, and smoking status (non-smoker, reformed smoker for more than 15 years, reformed smoker for less than or equal to 15 years, current smoker; coded as 0, 1, 2 and 3). These variables have been suggested in the literature (\cite{miller2004Bronchioloalveolar} and~\cite{nakachi1991genetic}). A total of 422 samples have clinical and environmental measurements available.

For the genetic part, we analyze gene expression data. A total of 18,969 measurements are available for 502 samples. When matching the clinical/environmental data with genetic data, complete data are available for 404 samples. Among them, 129 died during following. The median followup was 30 months. 275 were censored, and the median followup was 18 months.

The tuning parameter can be chosen using data-driven methods for example cross validation. However, we note that the commonly used methods are mostly based on the notion of prediction. With a large number of marginal models, the goal is to identify markers top-ranked in a marginal sense,
not prediction. Thus, following published studies, we vary the tuning so that a prefixed number of interactions are selected. In Table~\ref{tab3}, we provide results on the 33 top-ranked interactions. Longer or shorter lists of identified interactions are available from the authors.
After the interactions are identified, we refit the marginal models without penalization on the main effects to satisfy the ``main effects, interactions" hierarchy. Beyond the proposed method, we also apply the three alternatives considered in simulation. Table~\ref{tab7} in Appendix suggests that different methods identify significantly different genes and interactions. Results under the proposed method are provided in Table~\ref{tab3}. Those under the alternative methods are provided in Appendix.

To assess the stability of our findings, we apply a cross validation-based approach. Specifically, one sample is removed from data, and the proposed method is applied. This step is repeated over all samples. For each identified interaction in Table 3, we compute its probability of being identified in the reduced datasets. It can be seen that three interactions have very small stability measures. All other interactions have stability measures close to 1. We have also examined those interactions not identified and found that their stability measures are all close to 0. This analysis suggests a certain degree of stability of the proposed method.

Literature search suggests that the identified interactions and corresponding genes may have important implications. Specifically, previous studies \citep{shi2014gene} have suggested that the major G$\times$E interactions occur between genes and smoking status. Among the identified gene-smoking interactions in the current study, the CEBPB (CCAAT/enhancer-binding proteins beta) protein is a transcription factor that works with other CCAAT/enhancer-binding protein family members in the regulation of cell cycle progress, differentiation and pro-inflammatory gene expression. CEBPB has been found to be upregulated by tobacco smoke in both human lung fibroblast~\citep{Miglino2012Cigarette} and mice emphysema~\citep{hirama2007increased}. Another gene that interacts with smoking is
EFNA1 (a.k.a. ephrin-A1). A recent {\em in vitro} study found that ephrin-A1 is overexpressed in tobacco smoke treated bronchial airway epithelial cells compared to control cells~\citep{nasreen2014tobacco}. In addition, the elevated expressions of ephrin-A1 are positively associated with tumor proliferative capacity in non-small cell lung carcinoma patients~\citep{giaginis2014ephrin}. The gene that shows both a strong main effect and interaction with duration of smoking is TERF1 (telomeric repeat binding factor 1). The TERF1 expression levels have been found to be decreased in lung cancer (\cite{lin2006expression},~\cite{hu2009analysis}), as well as other types of cancer in several studies (\cite{yamada2002down},~\cite{miyachi2002correlation}). It functions as an inhibitor of telomerase and is identified as a prognostic marker for overall survival in non-small cell lung cancer~\citep{hu2009analysis}. The molecular mechanism of why and how TERF1 decreases in the process of cancer is not clear. Our results suggest that smoking can be one of the factors.
Other than smoking duration, we also find that several genes interact with smoking intensity as well. The gene that interacts with both gender and smoking intensity is KATNB1. KATNB1 encodes protein katanin p80 subunit B1, which has been found to participate in cytokinesis by interacting with tumor suppressor gene LAPSER1. The disruption of cytokinesis process may potentially cause genetic instability and cancer~\citep{Sudo2007LAPSER1}.
In addition to smoking, we find eight genes interact with gender in Table~\ref{tab3}. Among these, STRADB and CA5BP1 draw our attention. STRADB is an important gene in lung cancer progression and metastasis through the activation of LKB1. LKB1 is essential for G1 cell cycle arrest, cell polarity and stress, cell detachment and adhesion~\citep{marcus2010LKB1}. The STRADB encoded protein also interacts with the X chromosome-linked inhibitor of apoptosis protein by enhancing its anti-apoptotic activity. In addition, gene CA5BP1 is located on X chromosome and found to be gender-associated.

\section*{Discussion}
G$\times$E interactions have important implications for the prognosis of a large number of complex diseases. In this study, we have developed a novel interaction analysis method. It is robust to contamination in the prognosis outcome, which is not rare in practice but cannot be accommodated using most of the existing methods. In addition, we adopt penalization for identifying important interactions. This strategy differs from the commonly adopted significance level-based identification and can have better marker identification accuracy. Significantly advancing from most of the existing interaction studies, the consistency property has been rigorously established under the ultrahigh dimensional setting, making the proposed method one of the very few with a strong theoretical basis. An effective computational algorithm has been developed. In simulation study, the proposed method is observed to have satisfactory performance. With contamination, robust methods are needed. Under the majority of the cases, the proposed method outperforms the quantile regression based method. Thus, it provides a useful alternative to the quantile regression-based and other existing methods. In the analysis of lung cancer data, the proposed method identifies meaningful genes and interactions, which differ from those using the alternatives, with satisfactory stability.

We choose the popular Lasso penalty because of its low computational cost and satisfactory performance. A limitation of Lasso is that its results may not respect the ``main effects, interactions" hierarchy. However, as shown in Appendix, the proposed method can consistently identify the important interactions. As demonstrated in data analysis, once the important interactions are identified, if the hierarchy is desired, the main effects can be added back. With a robust loss function, the proposed method is computationally more expensive. However, as the computing can be executed in a highly parallel manner, and with the effective coordinate descent algorithm, the computer time is very much affordable. A total of 70 scenarios are considered in simulation and show the satisfactory performance of proposed method. More comprehensive comparisons of different methods, especially robust methods, have not been conducted in the literature and are of interest to future studies. Interesting findings are made in the lung cancer prognosis data analysis. Unfortunately there is still a lack of objective way of determining which set of results is more sensible. Validation of the findings will be pursued in future studies.


\clearpage
\begin{table}
	\centering
\caption{Simulation: AUC$\times$100 (sd) based on 100 replicates, $p=500$.}
	\begin{tabular}{ccccccc}
\hline
		Error&  Method     & Independent   & AR(0.2)   & AR(0.8)   & Band(0.3)  & Band(0.6)  \\
		\hline
N(0,1) & Robust & 86.1(7.7) & 90.1(6.6) & 86.7(6.9) & 89.2(5.7) & 94.3(5.8) \\
 & Unrobust & 88.9(6.6) & 88.1(5.7) & 86.8(6.1) & 83.6(6.3) & 91.0(8.2) \\
 & Stute & 65.7(4.4) & 63.3(3.3) & 61.6(4.6) & 64.7(3.5) & 61.0(2.3) \\
 & Quantile & 82.9(3.3) & 86.3(1.2) & 93.5(2.1) & 88.9(2.5) & 76.5(5.1) \\
 \hline
0.95N(0,1)+ & Robust & 91.1(7.8) & 87.1(6.9) & 88.7(5.1) & 87.7(6.6) &  88.1(6.5) \\
0.05Cauchy & Unrobust & 82.8(8.6) & 85.9(10) & 88.7(8.5) & 81.9(12.6) &78.5(10.2) \\
 & Stute & 68.0(4.7) & 60.1(4.6) & 70.6(4.8) & 64.7(3.5) & 73.3(4.7) \\
 & Quantile & 82.7(3.1) & 83.4(3.4) & 92.4(2.9) & 87.4(2.5) & 71.9(3.9) \\
 \hline
0.85N(0,1)+ & Robust & 86.6(8.2) &89.2(7.0) & 93.9(6.9) & 86.8(6.9) & 83.5(5.8) \\
0.15Cauchy & Unrobust & 77.1(9.4) & 85.0(10.2) & 87.1(12.7) & 72.8(10.3) & 76.2(11.6) \\
 & Stute & 64.2(6.3) & 63.7(5.5) & 67.5(6.9) & 64.7(3.5) & 67.8(6.4) \\
 & Quantile & 81.9(3.1) & 84.2(1.7) & 93.6(2.7) & 89.4(2.4) & 69.1(5.1) \\
 \hline
0.7N(0,1)+ & Robust & 84.8(6.9) & 88.6(6.6) & 87.9(5.9) & 89.2(5.7) & 90.1(5.9)\\
0.3Cauchy & Unrobust & 71.7(11.8) & 75.1(13.3) & 77.8(12.8) & 80.8(6.5) & 86.3(8.2) \\
 & Stute & 64.5(8.2) & 59.4(5.8) & 65.7(8.5) & 64.7(3.5) & 64.4(6.6) \\
 & Quantile & 80.1(2.7) & 85.8(2.3) & 92.5(2.4) & 88.2(2.9)& 68.1(8.3) \\
 \hline
0.95N(0,1)+ & Robust & 80.8(7.3) & 89.4(6.1)& 93.4(5.0) & 69.9(5.4) &88.8(6.7) \\
0.05t(3) & Unrobust & 76.7(9.1) & 84.7(8.4) & 89.6(9.0) & 82.1(11.6) & 85.4(10.1) \\
 & Stute & 60.8(4.5) & 66.7(4.9) & 68.6(5.0) & 64.7(3.5) & 73.8(4.9) \\
 & Quantile & 82.6(2.1) & 85.4(2.6) & 89.7(2.5) & 87.8(2.2) & 73.4(6.6) \\
 \hline
0.85N(0,1)+ & Robust & 87.5(7.5)& 85.6(6.6) & 90.6(6.0) & 87.9(6.5) &79.8(5.8) \\
0.15t(3) & Unrobust & 82.3(11.7) & 79.4(10.6) & 83.5(11.8) &75.5(13.9)&75.5(12.2) \\
 & Stute & 67.4(5.4) & 61.4(5.1) & 68.6(5.0) & 64.7(3.5) & 68.2(4.4) \\
 & Quantile & 83.1(3.4) & 85.3(3.1) &92.4(1.6)& 87.7(3.6) & 72.7(4.1) \\
 \hline
0.7N(0,1)+ & Robust & 84.4(7.2) & 88.6(6.8)& 86.3(6.7) & 88.4(5.3) &85.4(5.3) \\
0.3t(3) & Unrobust & 71.9(11.7) & 71.5(12.3) & 76.0(13.1) & 80.7(6.4) & 80.0(4.8) \\
 & Stute & 62.3(8.4) & 62.0(7.4) & 68.6(5.0) & 64.7(3.5) & 60.6(6.9) \\
 & Quantile & 79.8(4.1) & 83.9(2.6) & 93.3(1.9)& 87.1(3.1)& 68.5(6.3)\\ \hline
	\end{tabular}%
	\label{tab1}%
\end{table}%

\begin{table}
	\caption{Simulation: AUC$\times$100 (sd) based on 100 replicates, $p=1,000$.}
	\label{tab2}
	\centering
	\begin{tabular}{ccccccc}
\hline
		Error&  Method     & Independent & AR(0.2) & AR(0.8) & Band(0.3) & Band(0.6) \\
		\hline
		N(0,1)     & Robust & 89.2(6.8)& 87.4(6.7)&94.0(5.0)&92.8(7.2)& 85.4(5.3) \\
		& Unrobust & 84.4(6.1) &87.9(6.3) & 91.7(4.5) & 86.8(7.3) & 80.0(4.8) \\
		& Stute & 64.7(6.0) & 66.7(4.2) & 68.6(5.0) & 76.5(4.9) & 73.4(4.0) \\
		& Quantile & 82.4(2.2) & 85.1(3.1) & 92.8(4.2) & 88.7(2.7) & 71.1(3.9) \\
		\hline
		0.95N(0,1)+     & Robust &86.0(7.4)& 76.4(6.1) & 87.4(5.1) & 89.3(6.2)& 94.5(5.7)\\
		0.05Cauchy     & Unrobust & 82.0(8.9) & 83.9(10.6) & 88.4(5.5) & 80.0(11.6) & 91.2(7.2) \\
		& Stute & 62.2(4.6) & 64.6(5.9) & 68.6(5.0) & 65.4(4.1) & 71.5(4.6) \\
		& Quantile & 82.3(3.4) & 85.9(2.3)& 90.6(2.1) & 88.4(1.9) & 70.9(6.6) \\
		\hline
		0.85N(0,1)+     & Robust & 85.4(6.8) &93.4(7.2)& 88.3(7.2) & 88.1(5.5) & 92.4(6.2)\\
		0.15Cauchy     & Unrobust & 78.4(10.5) & 83.1(11.4) & 84.8(8.2) & 75.5(10.8) & 83.0(12.5) \\
		& Stute & 65.8(7.0) & 72.2(6.2) & 67.5(6.7) & 61.2(5.1) & 70.4(7.3) \\
		& Quantile & 82.1(3.1) & 85.7(1.9) & 92.5(2.3) & 87.6(3.1) & 69.3(5.8) \\
		\hline
		0.7N(0,1)+     & Robust &88.0(7.4)& 87.5(7.5) & 90.7(5.5) & 92.8(6.1) & 82.7(6.4)\\
		0.3Cauchy     & Unrobust & 72.7(10.4) & 74.4(14) & 89.3(7.5) & 92.5(8.9) & 74.4(10.5) \\
		& Stute & 58.2(6.2) & 64.9(7.8) & 68.6(5.0) & 63.0(4.9) & 62.4(6.6) \\
		& Quantile & 82.3(2.6) &86.3(2.7)& 92.4(3.1)& 88.6(2.1) & 70.6(3.4) \\
		\hline
		0.95N(0,1)+     & Robust & 73.4(5.8) & 87.7(6.5) & 92.5(5.1) & 78.8(5.6) & 90.6(6.3) \\
		0.05t(3)     & Unrobust & 78.9(7.5) &85.8(9.9) & 87.6(10.0) &  80.7(9.4) & 89.0(8.4) \\
		& Stute & 66.2(4.8) & 68.9(5.5) & 68.6(5.0) & 68.3(5.1) & 68.9(4.2) \\
		& Quantile & 80.4(3.1) & 83.1(4.1) & 89.9(2.4) &86.6(4.1) & 69.4(7.0) \\
		\hline
		0.85N(0,1)+     & Robust & 83.0(7.5) & 84.7(8.2)& 82.6(5.8) & 79.4(7.0) & 86.4(6.2)\\
		0.15t(3)     & Unrobust & 74.5(9.8) & 76.6(11.2) & 74.7(12.5) & 76.4(13.6) & 77.5(10.0) \\
		& Stute & 57.2(5.3) & 64.6(6.2) & 68.6(5.0) & 67.6(5.4) & 65.8(6.1) \\
		& Quantile & 81.6(3.8) & 83.9(3.3) & 82.3(2.4) &86.6(2.2) & 69.1(4.9) \\
		\hline
		0.7N(0,1)+     & Robust & 85.2(7.3)&90.0(6.7)& 85.8(5.6) &94.8(5.3)& 81.5(5.9)\\
		0.3t(3)     & Unrobust & 75.6(11.4) & 75.2(10.3) & 81.8(6.4) & 83.0(5.3) & 74.7(11.3) \\
		& Stute & 57.2(6.2) & 64.8(6.6) & 68.6(5.0) & 63.3(7.4) & 62.0(5.9) \\
		& Quantile & 79.7(2.7) & 83.6(1.6) & 91.9(2.3)& 87.9(1.9) & 68.3(6.1) \\
\hline
	\end{tabular}
\end{table}

\begin{sidewaystable}
	\centering
	\caption{Analysis of the lung cancer data using the robust method: estimates $\times 100$. For the interactions, values in ``()" are the stability results.}
	\label{tab3}
	\begin{tabular}{lccccc|cccc}
		\hline
		&\multicolumn{5}{c}{Main effects}&\multicolumn{4}{c}{Interactions}\\
		\hline
		& age & gender & intensity & status & gene & age & gender & intensity & status \\
		\hline
AP1S2 & -1.0 & -22.6 & -0.1 & 37.4 & 20.8 &   &   & -0.2(0.998) &   \\
BTD & 0.8 & -8.6 & 0.2 & 5.1 & -56.7 &   &   &   & 15.9(1.000) \\
C10ORF54 & 0.7 & 2.7 & 0.2 & 6.2 & 3.2 &   &   & -0.5(0.998) &   \\
CA5BP1 & -1.2 & -6.3 & 0.1 & 29.0 & 7.5 &   & 14.8(1.000) &   &   \\
CAPN1 & -3.5 & -8.9 & 0.3 & -8.6 & -50.0 &   & -24.7(0.000) &   &   \\
CEBPB & -1.0 & -15.5 & 0.0  & 29.4 & -41.6 &   &   &   & 8.1(0.998) \\
EFNA1 & -0.7 & 5.9 & 1.0 & 4.0 & -65.3 &   &   &   & 12.7(1.000) \\
FAM107B & -1.2 & -19.3 & -0.1 & 16.2 & 29.2 &   &   & -0.8(1.000) &   \\
FLRT3 & -1.2 & 24.6 & 0.8 & 6.1 & 33.4 &   &   & -1.5(0.998) &   \\
KATNB1 & -0.0 & 0.5 & 0.2 & 26.6 & -76.2 &   & 39.8(0.995) & 30.6(0.059) &   \\
LRRC1 & -3.6 & -13.6 & -0.3 & 64.5 & 25.5 &   & -24.3(0.995) &   &   \\
LYRM5 & 0.8 & -10.2 & 0.1 & 4.4 & -12.0 &   & -6.8(1.000) &   &   \\
AP1S2.1 & -1.0 & -22.6 & -0.1 & 37.4 & 20.8 &   &   & -0.2(0.998) &   \\
MYO18A & 1.5 & -15.9 & 0.1 & -7.8 & 16.3 &   &   & 0.2(1.000) &   \\
NOD1 & -1.3 & 16.0 & -0.1 & 4.6 & -28.9 &   &   & -0.3(0.998) &   \\
NPLOC4 & -3.3 & -34.8 & 0.7 & 58.4 & -76.3 &   &   &   & 14.4(0.998) \\
PLEKHO2 & 0.2 & 11.2 & 0.1 & 1.1 & -15.3 &   &   & -0.5(0.998) &   \\
POLR3GL & -4.3 & -1.7 & 0.0  & 63.0 & 28.2 &   &   & 0.1(0.998) &   \\
RAB27A & 0.6 & -12.9 & 0.0  & 10.1 & 14.4 &   &   & -0.5(0.998) &   \\
SECISBP2L & -3.7 & -34.7 & -0.5 & 55.3 & 30.4 &   &   & -0.8(0.998) &   \\
SGTB & 2.0 & -17.5 & 0.2 & 3.3 & 7.4 &   &   & 0.2(0.995) &   \\
STRADB & -0.7 & -3.4 & 0.1 & 4.1 & 1.2 &   & -48.5(0.995) &   &   \\
SWSAP1 & 1.9 & -14.6 & 0.1 & -2.9 & -12.1 &   &   &   & 0.8(0.995) \\
TEP1 & 1.7 & -33.3 & 0.3 & -0.9 & 5.4 &   &   & 0.3(1.000) &   \\
TERF1 & 0.2 & -22.0 & -0.0 & 43.2 & 46.4 &   &   &   & -18.9(1.000) \\
THOC1 & 0.9 & 30.9 & 1.2 & -0.8 & 31.4 &   & -67.4(0.002) &   &   \\
TIGD5 & 1.5 & -11.6 & 1.1 & -32.5 & 3.9 &   &   &   & -10.4(1.000) \\
TK2 & -1.0 & -13.8 & -0.4 & 25.0 & 24.7 &   &   & -0.8(1.000) &   \\
TMEM54 & 3.7 & -40.0 & 0.5 & -33.2 & -6.1 &   & -69.2(1.000) &   &   \\
TMEM106A & -0.8 & -11.2 & 0.2 & 20.4 & 4.1 &   &   & -0.3(0.998) &   \\
TOMM7 & 1.8 & 0.1 & 0.1 & -7.7 & -21.4 &   &   &   & 2.3(0.998) \\
TRIM34 & 0.4 & 5.5 & 0.2 & 2.1 & -13.4 &   &   & -0.3(0.998) &   \\
YARS2 & 0.5 & 8.1 & 0.9 & -9.7 & -5.8 &   &   &   & -2.5(0.998) \\
		\hline
	\end{tabular}
\end{sidewaystable}

\clearpage
\begin{figure}[htb]
\center{\includegraphics[width=0.7\textwidth, angle=0]{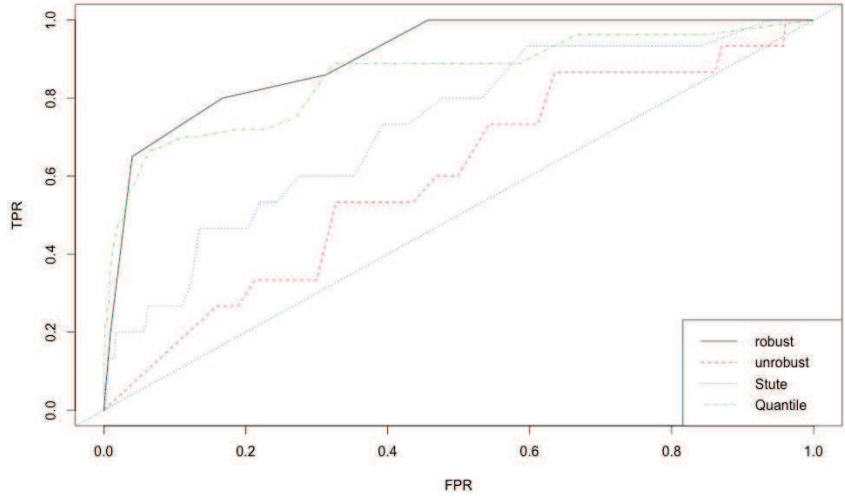}}
\caption{\label{fig:my-label} An illustration of the ROC curves for the proposed and alternative methods.}
\end{figure}

\clearpage\setcounter{page}{1}
\begin{center}
{\Large\bf Appendix}
\end{center}

\section*{Consistency properties}

Here we rigorously prove that the proposed method can consistently identify the important interactions under ultrahigh-dimensional settings. With the consistency properties, the proposed method can be preferred over the alternatives whose statistical properties have not been well established.

First we show that, under mild conditions, the proposed method can distinguish the important effects from unimportant ones in the presence of censoring. For each $j\in \{1,2,\cdots,p\}$, define the population version of the marginal estimate as
$$\bm\zeta_j^M =  {\arg\max}_{\bm\zeta_j} E\{\exp(-(T - \mathbf{U}_{j}^\top\bm\zeta_j)^2 / \theta)\},$$
where $E$ denotes expectation under the true model. Denote the $k$th element of $\bm{\zeta}_j$ by ${\zeta}_{j,k}$. The corresponding important covariate effect index set in $\bm{\zeta}_j^M$ is labeled as $S_j=\{k\in \{1,\ldots, 2q+2\}:\zeta_{j,k}^M \neq 0\}$. Denote $\mathcal {A}_X = \cup_{j=1}^p \{t: t\in S_j, 2\leq t\leq q+1\}$ as the important set with its corresponding environmental variables important in at least one marginal model. If $q+2\in S_j$, then the $j$th gene is associated with the disease outcome in a marginal sense. The set $\{t: t\in S_j, q+3\leq t\leq 2q+2\}$ contains important interactions between the $j$th gene and environmental variables.
Then we have the important gene set $\mathcal {A}_G = \cup_{j=1}^p \{j: q+2\in S_j\}$ and interaction set $\mathcal {A}_I = \cup_{j=1}^p \{(j-1)p+t-q-2: t\in S_j, q+3\leq t\leq 2q + 2\}$. Denote $S^c$ and $|S|$ as the complement and cardinality of set $S$ respectively. Denote the $k$th element of $\mathbf{u}_{i,j}$ by $u_{(i,j)_k}$, $k =1,\ldots,2q+2$.

For each $j\in \{1,2,\cdots,p\}$, define $$D_{n}({\bm\zeta}_j) = \sum_{i=1}^n \omega_i \exp(-(y_i - \mathbf{u}_{i,j}^\top{\bm\zeta}_{j})^2 / \theta)\frac{2(y_i - \mathbf{u}_{i,j}^\top{\bm\zeta}_{j})}{\theta}\mathbf{u}_{i,j}$$
and
$$I_{n}({\bm\zeta}_j) = \frac{2}{\theta}\sum_{i=1}^n \omega_i \exp(-(y_i - \mathbf{u}_{i,j}^\top{\bm\zeta}_{j})^2 / \theta)\left(\frac{2(y_i - \mathbf{u}_{i,j}^\top{\bm\zeta}_{j})^2}{\theta}-1\right)\mathbf{u}_{i,j}
\mathbf{u}_{i,j}^\top .$$
Let $\tau_Y, \tau_T$ and $\tau_C$ be the end points of the support of $Y,T$ and $C$, respectively. Assume the following regularity conditions:
\begin{description}
\item[C1.] The observations $\{(y_i, \delta_i, \mathbf{x}_i, \mathbf{z}_i), 1\leq i \leq n\}$ are independent and identically distributed;
  \item[C2.] $T$ and $C$ are independent and $P(T \leq C|T,X,Z) = P(T \leq C|T)$;
  \item[C3.] $\tau_T<\tau_C$ or $\tau_T= \tau_C=\infty$;
  \item[C4.] $q$ is a fixed. For each $j\in \{1,2,\cdots,p\}$, $\sqrt{n}D_{n}({\bm\zeta}_j^M)\rightarrow_d N(0, \Sigma_j)$, where $\Sigma_j$ is a positive definite matrix.
  $I_{n}({\bm\zeta}_j^M)$ converges to some negative definite matrix $I({\bm\zeta}_j^M)$ in probability. Moreover, the smallest eigenvalue $\rho_* = \min_{j}\rho_{\min} (-I({\bm\zeta}_j^M))$ and the largest eigenvalue $\rho^* = \max_{j}\rho_{\max} (\Sigma_j)$ are bound away from zero and infinity.
  \item[C5.] Let $N_j$ denote a sufficient small neighborhood centered at ${\bm\zeta}_j^M$. For ${\bm\zeta}_j^1, {\bm\zeta}_j^2\in N_j$, there exists a bounded constant $V$ such that ${\bm\zeta}_j^\top [I({\bm\zeta}_j^1)-I({\bm\zeta}_j^2)]{\bm\zeta}_j \leq V \|{\bm\zeta}_j^1-{\bm\zeta}_j^2\|_2$ with any $\|{\bm\zeta}_j\|_2 = 1$. Moreover, For any $k_1, k_2\in \{1,\ldots, 2q+2\}$, $E(\exp(-(y_i - \mathbf{u}_{i,j}^\top{\bm\zeta}_{j}^*)^2 / \theta)\left(\frac{2(y_i - \mathbf{u}_{i,j}^\top{\bm\zeta}_{j}^*)^2}{\theta}-1\right)u_{(i,j)_{k_1}}u_{(i,j)_{k_2}})^2 \leq J$, where $J$ is finite and ${\bm\zeta}_{j}^*\in N_j$.
\end{description}
\begin{remark}
C1-C3 have been commonly assumed in models with random censoring. Condition C4 is mild and has been proved under some regular conditions in ~\cite{Stute1993consistent} and \cite{huang2007least}. C5 is added to simplify the proof.
\end{remark}

If the truly important effects were known, then we would be able to compute the oracle estimator. Consider the oracle estimation $\hat{\bm\zeta}_j$ with $\hat{\bm\zeta}_{S_j^c}=0$ and
\begin{equation}\label{oracle}
\hat{\bm\zeta}_{S_j} = {\arg\max} \sum_{i=1}^n\omega_i\exp(-(y_i - \mathbf{u}_{(i,j)_{S_j}}^\top{\bm\zeta}_{j,S_j})^2 / \theta)-\lambda\sum_{k\in S_j}|\zeta_{j,k}|.
\end{equation}

\begin{theorem}\label{11}
Consider the estimator defined in (\ref{oracle}). Under Condition C1-C5 and $6(n^{-\kappa}+ 12\rho_*^{-1} (q+1) \lambda)V < \rho_*$, we have
\begin{eqnarray*}
&&\Pr\left(\max_{1\leq j \leq p} \max_{s\in S_j}|\hat\zeta_{j,s}-{\zeta}_{j,s}^M|\geq n^{-\kappa}+12 q \rho_*^{-1} \lambda \right)\\
&&~~~\leq p\exp\left(-\frac{\rho_*^2 n^{1-2\kappa} + 144 q^2 n\lambda^2}{36  \rho^* }\right)+4p(q+1) \exp\left(-\frac{n \rho_*^2}{36J(q+1)^2}\right),
\end{eqnarray*}
where $\kappa< 1/2$.
\end{theorem}
The tail probability in Theorem \ref{11} is exponentially small. In other words, the proposed method is able to accommodate ultrahigh dimensional data with
\[\log p = o (n^{1-2\kappa}+ n\lambda^2).\]

Recall that $\tilde{\bm\zeta}_j = {\arg\max}_{\bm\zeta\in \mathbb{R}^{2q+2}}L(\bm\zeta_j)$, where
\begin{equation}\label{app1}
L(\bm\zeta_j) = \sum_{i=1}^n\omega_i\exp(-(y_i - \mathbf{u}_{i,j}^\top\bm\zeta_j)^2 / \theta)-\lambda||\bm\zeta_j||_1.
\end{equation}
Since $L(\bm\zeta_j)$ in (\ref{app1}) is concave, if we can show the oracle estimator $\hat{\bm\zeta}_j$ satisfies the Karush-Kuhn-Tucher(KKT) condition, then  $\tilde{\bm\zeta}_j= \hat{\bm\zeta}_j$.

Define $\widetilde{\mathcal{A}_G}=\cup_{j=1}^p \{j: \tilde{\zeta}_{j, q+2}\neq0\}$ and $\widetilde{\mathcal{A}_I} = \cup_{j=1}^p \{(j-1)p+t-q-2: \tilde{\zeta}_{j, t}\neq0, q+3\leq t\leq 2q + 2\}$. Partition $I({\bm\zeta}_j^M)$ according to $S_j$ as
\[I({\bm\zeta}_j^M) = \left(
                        \begin{array}{cc}
                          I_{S_j S_j}({\bm\zeta}_j^M) & I_{ S_j {S_j}^c}({\bm\zeta}_j^M) \\
                          I_{{S_j}^c  S_j }({\bm\zeta}_j^M) & I_{{S_j}^c {S_j}^c}({\bm\zeta}_j^M) \\
                        \end{array}
                      \right).
\]
The following theorem establishes that the proposed method has the asymptotic consistency properties.
\begin{theorem}\label{22}
Assume Condition C1-C4, $\kappa< 1/2$ and $\Phi_j= \|I_{{S_j}^c S_j}({\bm\zeta}_j^M) I_{{S_j}{S_j}}({\bm\zeta}_j^M)^{-1}\|_\infty \leq K<1$. If $\mathcal {O}_G\subseteq \mathcal {A}_G $ and $\mathcal {O}_I\subseteq \mathcal {A}_I$, we have
$$\Pr\left( \mathcal {O}_G \subseteq \widetilde{\mathcal {A}_G}~ \mbox{and} ~\mathcal {O}_I\subseteq \widetilde{\mathcal {A}_I}\right)\geq 1-O\left(p\exp\left(-\frac{\rho_*^2 n^{1-2\kappa} + 144 q^2 n\lambda^2}{36  \rho^* }\right)+p\exp\left(-\frac{n \lambda^2(1-K)^2}{2 \rho^* (1+K)^2}\right)\right). $$
\end{theorem}

Below we provide proofs for the two theorems.

\noindent{\bf Proof of Theorem~\ref{11}}

Recall that
\begin{equation}
\hat{\bm\zeta}_{j,S_j} = {\arg\max} \sum_{i=1}^n\omega_i\exp(-(y_i - \mathbf{u}_{(i,j)_{S_j}}^\top{\bm\zeta}_{j,S_j})^2 / \theta)-\lambda\sum_{k\in S_j}|\zeta_{j,k}|.
\end{equation}
Denote the above objective function as $R_n({\bm\zeta}_{j,S_j})$.

First, let $r_j = n^{-\kappa}+ 12\rho_*^{-1} (q+1) \lambda$ and $\eta = \sum\limits_{j=1}^p \exp\left(-\frac{\rho_*^2 n^{1-2\kappa} + 144 (q+1)^2 n^{-1}\lambda^2}{36  {\rho^*} }\right)$, where $\kappa< 1/2$. To prove that
\[\Pr\left\{\|{\bm\zeta}_{j,S_j} - {\bm\zeta}_{j,S_j}^M\|_2 < r_j, j=1,\cdots,p\right\}\geq 1-\eta,\]
it suffices to show that
\begin{equation}
\Pr\Big(\sup\limits_{
{\bm\zeta}_{j,S_j}\in \mathcal{I}}R_n({\bm\zeta}_{j,S_j})<R_n({\bm\zeta}_{j,S_j}^M), j=1,\cdots,p \Big)\geq 1-\eta,
\end{equation}
where $\mathcal{I}=\left\{{\bm\zeta}_{j,S_j}: \|{\bm\zeta}_{j,S_j} - {\bm\zeta}_{j,S_j}^M\|_2 = r_j, j=1,\cdots,p\right\}$. This implies that with probability at least $1-\eta$, $R_n({\bm\zeta}_{j,S_j})$ has a global
maximizer $\hat {\bm\zeta}_{j,S_j}$ that satisfies
$\|{\bm\zeta}_{j,S_j} - {\bm\zeta}_{j,S_j}^M\|_2 < r_j$, for $j=1,\cdots,p$.

Recall the definitions of $D_{n}({\bm\zeta}_{j})$ and $I_{n}({\bm\zeta}_{j})$. Partition them according to $S_j$ as
\[D_{n}({\bm\zeta}_{j})=\left(
                          \begin{array}{c}
                            D_{n, S_j}({\bm\zeta}_{j}) \\
                            D_{n, S_j^c}({\bm\zeta}_{j}) \\
                          \end{array}
                        \right), ~~I_{n}({\bm\zeta}_{j})=\left(
                                                         \begin{array}{cc}
                                                           I_{n, S_j S_j}({\bm\zeta}_{j}) & I_{n, S_j S_j^c}({\bm\zeta}_{j}) \\
                                                           I_{n, S_j^c S_j}({\bm\zeta}_{j}) & I_{n, S_j^c S_j^c}({\bm\zeta}_{j}) \\
                                                         \end{array}
                                                       \right).
\]
Obviously, $$D_{n,S_j}({\bm\zeta}_{j}) = \sum_{i=1}^n \omega_i \exp(-(y_i - \mathbf{u}_{(i,j)_{S_j}}^\top{\bm\zeta}_{j,S_j})^2 / \theta)\frac{2(y_i - \mathbf{u}_{(i,j)_{S_j}}^\top{\bm\zeta}_{j,S_j})}{\theta}\mathbf{u}_{(i,j)_{S_j}}$$
and
 $$I_{n, {S_j}{S_j}}({\bm\zeta}_{j}) = \frac{2}{\theta}\sum_{i=1}^n \omega_i \exp(-(y_i - \mathbf{u}_{(i,j)_{S_j}}^\top{\bm\zeta}_{j,S_j})^2 / \theta)\left(\frac{2(y_i - \mathbf{u}_{(i,j)_{S_j}}^\top{\bm\zeta}_{j,S_j})^2}{\theta}-1\right)
 \mathbf{u}_{(i,j)_{S_j}}\mathbf{u}_{(i,j)_{S_j}}^\top .$$
In fact, by Taylor's expansion we have
\begin{eqnarray}\label{q0}
&&R_n({\bm\zeta}_{j,S_j})-R_n({\bm\zeta}_{j,S_j}^M)\nonumber\\
&=& \sum_{i=1}^n\omega_i\left\{\exp(-(y_i - \mathbf{u}_{(i,j)_{S_j}}^\top{\bm\zeta}_{j,S_j})^2 / \theta)-\exp(-(y_i - \mathbf{u}_{(i,j)_{S_j}}^\top{\bm\zeta}_{j,S_j}^M)^2 / \theta)\right\}\nonumber\\
&& -\lambda\sum_{k\in S_j}\left(|\zeta_{j,k}|-|\zeta_{j,k}^M|\right\}\nonumber\\
&=& D_{n,S_j}({\bm\zeta}_{j}^M)^\top ({\bm\zeta}_{j,S_j} - {\bm\zeta}_{j,S_j}^M)+ \frac{1}{2}({\bm\zeta}_{j,S_j} - {\bm\zeta}_{j,S_j}^M)^\top I_{n,{S_j}{S_j}}(\bar{\bm\zeta}_{j})({\bm\zeta}_{j,S_j} - {\bm\zeta}_{j,S_j}^M)\nonumber\\
&&+\lambda\sum_{k\in S_j}\left(|\zeta_{j,k}^M|-|\zeta_{j,k}|\right\},
\end{eqnarray}
where $\bar{\bm\zeta}_{j}$ lies between ${\bm\zeta}_{j}^M$ and ${\bm\zeta}_{j}$.

It is easy to see that
\begin{eqnarray}\label{q1-0}
&&({\bm\zeta}_{j,S_j} - {\bm\zeta}_{j,S_j}^M)^\top I_{n,{S_j}{S_j}}(\bar{\bm\zeta}_{j})({\bm\zeta}_{j,S_j} - {\bm\zeta}_{j,S_j}^M)\nonumber\\
&=&({\bm\zeta}_{j,S_j} - {\bm\zeta}_{j,S_j}^M)^\top I_{{S_j}{S_j}}({\bm\zeta}_j^M)({\bm\zeta}_{j,S_j} - {\bm\zeta}_{j,S_j}^M)\nonumber \\
&& + ({\bm\zeta}_{j,S_j} - {\bm\zeta}_{j,S_j}^M)^\top \left\{I_{{S_j}{S_j}}(\bar{{\bm\zeta}}_j)- I_{{S_j}{S_j}}({\bm\zeta}_j^M)\right\}({\bm\zeta}_{j,S_j} - {\bm\zeta}_{j,S_j}^M)\nonumber\\
&& + ({\bm\zeta}_{j,S_j} - {\bm\zeta}_{j,S_j}^M)^\top \left\{I_{n,{S_j}{S_j}}(\bar{\bm\zeta}_{j})-I_{{S_j}{S_j}}(\bar{{\bm\zeta}}_j)\right\}
({\bm\zeta}_{j,S_j} - {\bm\zeta}_{j,S_j}^M)\nonumber\\
&= & Q_1 + Q_2 + Q_3.
\end{eqnarray}
By C4, $Q_1\leq -\|{\bm\zeta}_{j,S_j} - {\bm\zeta}_{j,S_j}^M\|_2^2\rho_*$. Moreover, $Q_2\leq V\|{\bm\zeta}_{j,S_j} - {\bm\zeta}_{j,S_j}^M\|_2^3$ under C5. Bernstein inequality and C5 yield
\[\Pr(\|I_{n,{S_j}{S_j}}(\bar{\bm\zeta}_{j})-I_{{S_j}{S_j}}(\bar{{\bm\zeta}}_j)\|_F^2 \geq \frac{\rho_*^2}{9} ) \leq 2|S_j| \exp\left(-\frac{n \rho_*^2}{9J|S_j|^2}\right),\]
where $\|\cdot\|_F$ denotes the Frobenius norm. Since $\lambda_{\max}(I_{n,{S_j}{S_j}}(\bar{\bm\zeta}_{j})-I_{{S_j}{S_j}}(\bar{{\bm\zeta}}_j))\leq \|I_{n,{S_j}{S_j}}(\bar{\bm\zeta}_{j})-I_{{S_j}{S_j}}(\bar{{\bm\zeta}}_j)\|_F$, we have $Q_3\leq \frac{1}{3}  \rho_* r_j^2$.
Therefore, the second term in (\ref{q0}) can be controlled by
\begin{eqnarray}\label{q1}
 \frac{1}{2}({\bm\zeta}_{j,S_j} - {\bm\zeta}_{j,S_j}^M)^\top I_{n,{S_j}{S_j}}({\bm\zeta}_{j}^M)({\bm\zeta}_{j,S_j} - {\bm\zeta}_{j,S_j}^M)<  -\frac{1}{3}  \rho_* r_j^2 +\frac{1}{2}V r_j^3,
\end{eqnarray}
with probability at least $1-4(q+1) \exp\left(-\frac{n \rho_*^2}{36J(q+1)^2}\right)$ due to (\ref{q1-0}) and $|S_j|\leq 2q+2$.

Partition $\Sigma_j$ according to $S_j$ as $\left(
                                            \begin{array}{cc}
                                              \Sigma_{S_j S_j}&  \Sigma_{S_j S_j^c} \\
                                               \Sigma_{S_j^c S_j} &  \Sigma_{S_j^c S_j^c} \\
                                            \end{array}
                                          \right)
$. For $D_{n,S_j}({\bm\zeta}_j^M)$, by the definition of ${\bm\zeta}_{j,S_j}^M$, and conditions C2 and C4, we have
\begin{equation*}
\sqrt{n}D_{n,S_j}({\bm\zeta}_j^M)\rightarrow_d N(0, \Sigma_{S_j S_j}).
\end{equation*}
Then for any given $t$, an application of Bernstein's inequality,
\begin{eqnarray*}
\Pr(|D_{n,S_j}({\bm\zeta}_j^M)^\top ({\bm\zeta}_{j,S_j} - {\bm\zeta}_{j,S_j}^M)|>t )\leq 2\exp\left(-\frac{n t^2}{2 ({\bm\zeta}_{j,S_j} - {\bm\zeta}_{j,S_j}^M)^\top\Sigma_{S_j S_j}({\bm\zeta}_{j,S_j} - {\bm\zeta}_{j,S_j}^M)}\right).
\end{eqnarray*}
 Recall that $r_j = n^{-\kappa}+ 12\rho_*^{-1} (q+1) \lambda$. Let $t = \frac{1}{6}  \rho_* r_j^2$, then we have
\begin{eqnarray}\label{q2}
\Pr(D_{n,S_j}({\bm\zeta}_j^M)^\top ({\bm\zeta}_{j,S_j} - {\bm\zeta}_{j,S_j}^M)>\frac{1}{6} \rho_* r_j^2)\leq \exp\left(-\frac{\rho_*^2 n^{1-2\kappa} + 144 (q+1)^2 n \lambda^2}{36  {\rho^*} }\right).
\end{eqnarray}
By the Triangle inequality and $(\sum\limits_{i=1}^d |v_i|)^2 \leq d\sum\limits_{i=1}^d v_i^2$ for any sequence ${v_i}$, we have
\begin{eqnarray}\label{q3}
\lambda\sum_{k\in S_j}\left(|\zeta_{j,k}^M|-|\zeta_{j,k}|\right\} \leq \lambda\sum_{k\in S_j}|\zeta_{j,k}^M-\zeta_{j,k}|\leq \lambda\sqrt{|S_j|}\|{\bm\zeta}_{j,S_j} - {\bm\zeta}_{j,S_j}^M\|_2.
\end{eqnarray}
Combining (\ref{q0}), (\ref{q1}), (\ref{q2}),(\ref{q3}), and $6(n^{-\kappa}+ 12\rho_*^{-1} (q+1) \lambda)V < \rho_*$, we have
\begin{eqnarray}
R_n({\bm\zeta}_{j,S_j})-R_n({\bm\zeta}_{j,S_j}^M)< -\frac{ 1}{6} \rho_* r_j^2+ \lambda \sqrt{|S_j|}r_j+\frac{ 1}{2} V r_j^3 <0
\end{eqnarray}
with probability at least
$1-\exp\left(-\frac{\rho_*^2 n^{1-2\kappa} + 144 (q+1)^2 n\lambda^2}{36  {\rho^*} }\right)-4(q+1) \exp\left(-\frac{n \rho_*^2}{36J(q+1)^2}\right)$.
Together with the Bonferroni's inequality, we have the conclusion.  \hfill $\Box$

\noindent{\bf Proof of Theorem~\ref{22}}

Recall that $\tilde{\bm\zeta}_j = {\arg\max}_{\bm\zeta\in \mathbb{R}^{2q+2}}L(\bm\zeta_j)$, where
\begin{equation}\label{app11}
L(\bm\zeta_j) = \sum_{i=1}^n\omega_i\exp(-(y_i - \mathbf{u}_{i,j}^\top\bm\zeta_j)^2 / \theta)-\lambda||\bm\zeta_j||_1.
\end{equation}
Consider the oracle estimation $\hat{\bm\zeta}_j$ with $\hat{\bm\zeta}_{S_j^c}=0$ and
\begin{equation}
\hat{\bm\zeta}_{j,S_j} = {\arg\max} \sum_{i=1}^n\omega_i\exp(-(y_i - \mathbf{u}_{(i,j)_{S_j}}^\top{\bm\zeta}_{j,S_j})^2 / \theta)-\lambda\sum_{k\in S_j}|\zeta_{j,k}|.
\end{equation}
Denote the above objective function as $R_{n,S_j}({\bm\zeta}_{j}) $. Since $L(\bm\zeta_j)$ in (\ref{app1}) is concave, if we can show that the oracle estimation $\hat{\bm\zeta}_j$ satisfies Karush-Kuhn-Tucher(KKT) condition, then  $\tilde{\bm\zeta}_j= \hat{\bm\zeta}_j$. Hereafter we will focus on proving that the oracle estimation $\hat{\bm\zeta}_j$ satisfies KKT condition.

Next we want to show that
\begin{equation}\label{kkt2}
\left|\Omega_n(S_j^c)\right|_\infty<\lambda, ~~j=1, 2,\cdots, p
\end{equation}
where $|\nu|_\infty = \max_i |\nu_i|$ for any vector $\nu= (\nu_1, \cdots, \nu_{|S_j^c|})$ and
$$\Omega_n(S_j^c)= \sum_{i=1}^n \omega_i \exp(-\frac{(y_i - \mathbf{u}_{(i,j)_{S_j}}^\top\hat{\bm\zeta}_{j,S_j})^2 }{\theta} )\frac{2(y_i - \mathbf{u}_{(i,j)_{S_j}}^\top\hat{\bm\zeta}_{j,S_j})}{\theta}\mathbf{u}_{(i,j)_{S_j^c}}.$$
Applying Taylor's expansion, we have
\begin{eqnarray}\label{omega}
\Omega_n(S_j^c)&=& \sum_{i=1}^n \omega_i \exp\left\{-\frac{(y_i - \mathbf{u}_{(i,j)_{S_j}}^\top{\bm\zeta}_{j,S_j}^M)^2 }{\theta} \right\}\frac{2(y_i - \mathbf{u}_{(i,j)_{S_j}}^\top{\bm\zeta}_{j,S_j}^M)}
{\theta}\mathbf{u}_{(i,j)_{S_j^c}}\nonumber\\
&& + \frac{2}{\theta}\sum_{i=1}^n \omega_i \exp\left\{-\frac{(y_i - \mathbf{u}_{(i,j)_{S_j}}^\top\bar{{\bm\zeta}}_{j,S_j})^2 }{\theta}\right\}\left(\frac{2(y_i - \mathbf{u}_{(i,j)_{S_j}}^\top\bar{{\bm\zeta}}_{j,S_j})^2}{\theta}-1\right)\nonumber\\
&&~~~\times \mathbf{u}_{(i,j)_{S_j^c}}\mathbf{u}_{(i,j)_{S_j}}^\top (\hat{\bm\zeta}_{j,S_j} - {\bm\zeta}_{j,S_j}^M )\nonumber\\
&:= & \Gamma_{n} + \Delta_n,
\end{eqnarray}
where $\bar{\bm\zeta}_{j}$ lies between ${\bm\zeta}_{j}^M$ and $\hat{{\bm\zeta}}_{j}$. From the proof of Theorem 1, we have
\begin{eqnarray}\label{hatzeta}
\hat{\bm\zeta}_{j,S_j} - {\bm\zeta}_{j,S_j}^M = I_{n, {S_j}{S_j}}({\bm\zeta}_j^M)^{-1} \{-D_{n,S_j}({\bm\zeta}_j^M)+ \lambda \mbox{sgn} ({\bm\zeta}_j^M)\}.
\end{eqnarray}
Hence substituting (\ref{hatzeta}) into (\ref{omega}), we obtain
\begin{eqnarray}\label{delta}
\Delta_n &=&
 \frac{2}{\theta}\sum_{i=1}^n \omega_i \exp\left\{-\frac{(y_i - \mathbf{u}_{(i,j)_{S_j}}^\top\bar{\bm\zeta}_{j,S_j})^2 }{\theta}\right\}\left(\frac{2(y_i - \mathbf{u}_{(i,j)_{S_j}}^\top\bar{\bm\zeta}_{j,S_j})^2}{\theta}-1\right)\nonumber\\
&&~~~\times \mathbf{u}_{(i,j)_{S_j^c}}\mathbf{u}_{(i,j)_{S_j}}^\top I_{n, {S_j}}({\bm\zeta}_j^M)^{-1} \{-D_{n,S_j}({\bm\zeta}_j^M)+ \lambda \mbox{sgn} ({\bm\zeta}_j^M)\}\nonumber\\
&=& -I_{n, {S_j}^c S_j}(\bar{\bm\zeta}_{j}) I_{n, {S_j}{S_j}}({\bm\zeta}_j^M)^{-1}D_{n,S_j}({\bm\zeta}_j^M)+\lambda I_{n, {S_j}^c S_j}(\bar{\bm\zeta}_{j}) I_{n, {S_j}}({\bm\zeta}_j^M)^{-1}\mbox{sgn} ({\bm\zeta}_j^M).
\end{eqnarray}

Next we define
\begin{eqnarray*}
\Delta_n^*
&=& -I_{{S_j}^c S_j}({\bm\zeta}_{j}^M) I_{{S_j}{S_j}}({\bm\zeta}_j^M)^{-1}D_{n,S_j}({\bm\zeta}_j^M)+\lambda I_{{S_j}^c S_j}({\bm\zeta}_{j}^M) I_{ {S_j}{S_j}}({\bm\zeta}_j^M)^{-1}\mbox{sgn} ({\bm\zeta}_j^M),
\end{eqnarray*}
and $\Omega_n^*(S_j^c) =  \Gamma_{n} + \Delta_n^*$.
From the proof of Theorem 1, we find the tail probability for $I_n({\bm\zeta}_j)$ is dominated by that for $D_{n}({\bm\zeta}_j)$. Thus
$$\Pr(\left|\Omega_n(S_j^c)\right|_\infty>\lambda)\asymp \Pr(\left|\Omega_n^*(S_j^c)\right|_\infty>\lambda).$$
Therefore, combining the above discussion, we only need focus on $\Omega_n^*(S_j^c)$. In fact,
\begin{eqnarray}\label{ome}
&&|\Omega_n^*(S_j^c)|_\infty \leq  |\Gamma_{n}|_\infty+|\Delta_n^*|_\infty\nonumber\\
&\leq & |D_{n,S_j^c}({\bm\zeta}_j^M)|_\infty+|I_{{S_j}^c S_j}({\bm\zeta}_j^M) I_{{S_j}{S_j}}({\bm\zeta}_j^M)^{-1}D_{n,S_j}({\bm\zeta}_j^M)|_\infty\nonumber\\
&& ~+\lambda|I_{{S_j}^c S_j}({\bm\zeta}_j^M) I_{{S_j}{S_j}}({\bm\zeta}_j^M)^{-1}D_{n,S_j}({\bm\zeta}_j^M)\mbox{sgn} ({\bm\zeta}_j^M)|_\infty\nonumber\\
&\leq & |D_{n}({\bm\zeta}_j^M)|_\infty + \|I_{{S_j}^c S_j}({\bm\zeta}_j^M) I_{{S_j}{S_j}}({\bm\zeta}_j^M)^{-1}\|_\infty |D_{n}({\bm\zeta}_j^M)|_\infty+ \lambda  \|I_{{S_j}^c S_j}({\bm\zeta}_j^M) I_{{S_j}{S_j}}({\bm\zeta}_j^M)^{-1}\|_\infty
\end{eqnarray}

By the condition $\Phi_j= \|I_{{S_j}^c S_j}({\bm\zeta}_j^M) I_{{S_j}{S_j}}({\bm\zeta}_j^M)^{-1}\|_\infty \leq K<1$, if
\begin{eqnarray}\label{pb}
|D_{n}^*({\bm\zeta}_j^M)|_\infty <\lambda \frac{1-\Phi_j}{1+\Phi_j},
\end{eqnarray}
then from (\ref{ome}), it follows
\begin{eqnarray*}
|\Omega_n^*(S_j^c)|_\infty &\leq& |D_{n}({\bm\zeta}_j^M)|_\infty(1+\Phi_j)+ \lambda \Phi_j \\
&<& \lambda (1-\Phi_j) + \lambda \Phi_j=\lambda,
\end{eqnarray*}
which proves (\ref{kkt2}). We now derive the probability bounds for the event in (\ref{pb}). Similarly as the derivation of (\ref{q2}),
\begin{eqnarray}\label{tail1}
\Pr\left\{|D_{n}({\bm\zeta}_j^M)|_\infty \geq \lambda \frac{1-\Phi_j}{1+\Phi_j}\right\}\leq 2\exp\left(-\frac{n \lambda^2(1-\Phi_j)^2}{2 {\rho^*} (1+\Phi_j)^2}\right).
\end{eqnarray}
By the Bonferroni's inequality, we obtain
\begin{eqnarray}\label{tail}
\Pr\left\{|D_{n}({\bm\zeta}_j^M)|_\infty < \lambda \frac{1-\Phi_j}{1+\Phi_j}, j =1, 2,\cdots, p \right\}\geq 1- 2\sum_{j=1}^p\exp\left(-\frac{n \lambda^2(1-\Phi_j)^2}{2 {\rho^*} (1+\Phi_j)^2}\right).
\end{eqnarray}
Based on the above result, the theorem is proved.
\hfill $\Box$

\clearpage
\section*{Additional numerical results}

\begin{table}[h]
\centering
\caption{Summary analysis results for the lung cancer data using different methods. Diagonals are the numbers of identified genes using different methods. Off-diagonals are the numbers of overlapping genes. In ``()" are the numbers of overlapping interactions.}
\label{tab7}
\begin{tabular}{lllll}
\hline
& Robust & Unrobust & Stute & Quantile\\
Robust & 33 & 9(5) & 0(0) & 0(0)\\
Unrobust & - & 31 & 3(0) & 1(0)\\
Stute & - & - & 30 & 2(1)\\
quantile & - & - & - & 28\\
\hline
\end{tabular}
\end{table}

\begin{sidewaystable}
	\centering
	\caption{Analysis of the lung cancer data using the unrobust method: estimates $\times$100. }
	\label{tab4}
	\begin{tabular}{lccccc|cccc}
		\hline
		&\multicolumn{5}{c}{Main effect}&\multicolumn{4}{c}{Interaction}\\
		\hline
		& age & gender & intensity & status & gene & age & gender & intensity & status \\
		\hline
BCL10 & 5.5 & 79.8 & 0.3 & 105.6 & 10.3 &     &     &     & -29.2 \\
BTD & 6.1 & 38.4 & 0.9 & 91.5 & -187.4 &     &     &     & 94.8 \\
CAPN1 & 6.1 & 77.3 & -0.0 & 97.9 & 22.1 &     &     &     & -36.2 \\
CASK & 6.4 & 59.1 & 0.4 & 81.9 & 90.5 &     &     &     & -63.1 \\
CSNK1G2 & 5.9 & 31.0 & 0.5 & 98.0 & -20.5 &     & -70.4 &     &     \\
DOCK6 & 5.7 & 100.0 & 0.4 & 85.6 & 6.5 &     &     &     & -35.4 \\
ECI2 & 7.3 & 42.0 & 1.1 & 39.6 & -87.0 &     &     &     & 57.3 \\
ELMO3 & 5.1 & 88.5 & 0.1 & 122.5 & -34.5 &     &     &     & -12.1 \\
FAM83H & 5.6 & 57.3 & 0.5 & 116.1 & 113.1 &     &     &     & -72.0 \\
FASN & 6.0 & 53.2 & 0.7 & 93.0 & 117.7 &     &     &     & -78.4 \\
KATNB1 & 6.2 & 59.6 & 0.4 & 85.0 & 55.6 &     &     &     & -49.1 \\
LYRM5 & 7.4 & 35.4 & 0.9 & 40.4 & -43.4 &     & 12.2 & 32.8 &     \\
MACROD1 & 5.8 & 73.2 & 0.3 & 99.1 & 73.1 &     &     &     & -58.4 \\
NACC2 & 6.1 & 65.9 & 0.3 & 92.6 & 100.7 &     &     &     & -70.9 \\
PKP3 & 5.8 & 73.9 & 0.1 & 99.0 & 43.3 &     &     &     & -54.8 \\
RBFA & 5.8 & 62.4 & 0.1 & 104.4 & 10.9 &     & -66.1 &     &     \\
RNH1 & 6.0 & 70.6 & 0.5 & 79.8 & 14.9 &     &     &     & -32.7 \\
SCYL1 & 5.5 & 84.1 & 0.4 & 105.6 & -36.2 &     &     &     & -6.3 \\
STRADB & 6.1 & 59.7 & 0.9 & 76.1 & -108.6 &     &     &     & 62.6 \\
SWSAP1 & 5.8 & 69.7 & 0.4 & 92.9 & 18.0 &     &     &     & -30.6 \\
TEN1 & 5.6 & 66.3 & 0.6 & 104.1 & 21.9 &     &     &     & -34.3 \\
TIGD5 & 5.5 & 46.3 & 0.8 & 113.7 & 125.3 &     &     &     & -73.9 \\
TMEM54 & 6.1 & 54.6 & -0.0 & 101.4 & -1.3 &     & -81.0 &     &     \\
TNIP2 & 5.1 & 85.9 & 0.6 & 100.9 & -50.6 &     &     &     & -4.2 \\
TOLLIP & 6.3 & 68.2 & 0.5 & 73.2 & 39.8 &     &     &     & -48.4 \\
TTC22 & 5.6 & 72.3 & 0.1 & 114.0 & -20.3 &     & -51.8 &     &     \\
WASH2P & 5.7 & 25.2 & 0.3 & 109.1 & 19.9 &     & -129.7 &     &     \\
YARS2 & 8.1 & 18.8 & 1.1 & 13.9 & -41.6 &     &     &     & 26.5 \\
YIPF2 & 5.7 & 75.2 & 0.4 & 99.8 & 7.8 &     &     &     & -27.4 \\
ZNF512B & 6.3 & 54.7 & 0.4 & 89.8 & 94.9 &     & -45.3 & -49.6 &     \\
ZNF699 & 5.8 & 93.7 & 0.5 & 86.0 & 59.1 &     &     &     & -56.4 \\
		\hline
\end{tabular}
\end{sidewaystable}

\begin{sidewaystable}
	\centering
	\caption{Analysis of the lung cancer data using the Stute method: estimates $\times$100.}
	\label{tab5}
	\begin{tabular}{lccccc|cccc}
		\hline
		&\multicolumn{5}{c}{Main effects}&\multicolumn{4}{c}{Interactions}\\
		\hline
		& age & gender & intensity & status & gene & age & gender & intensity & status \\
		\hline
KLHL9 & 8.1 & 17.9 & 1.0 & 29.2 & -358.5 & 5.4 &     &     &     \\
TPD52L2 & 7.5 & 56.8 & 1.0 & 31.2 & 313.5 & -4.7 &     &     &     \\
TMEM129 & 7.7 & 21.9 & 1.1 & 28.5 & -522.3 & 7.4 &     &     &     \\
PCGF3 & 8.2 & 20.2 & 1.1 & 15.9 & -527.4 & 7.6 &     &     &     \\
XPNPEP1 & 7.8 & 30.0 & 0.9 & 26.1 & -544.6 & 7.9 &     &     &     \\
FDPS & 7.6 & 24.8 & 1.1 & 37.0 & 254.2 & -3.8 &     &     &     \\
GFOD2 & 7.3 & 24.7 & 1.0 & 40.8 & -447.1 & 6.1 &     &     &     \\
MAGED4B & 7.3 & 23.3 & 1.0 & 44.8 & -654.5 & 9.2 &     &     &     \\
PIGG & 8.0 & 11.6 & 1.0 & 27.6 & -354.2 & 5.1 &     &     &     \\
UVSSA & 8.3 & 21.3 & 1.0 & 17.1 & -527.7 & 7.7 &     &     &     \\
LAMTOR2 & 7.4 & 37.0 & 1.0 & 36.8 & 263.9 & -3.7 &     &     &     \\
CSNK1G2 & 7.2 & 26.4 & 0.9 & 44.5 & -390.9 & 5.2 &     &     &     \\
POLR3C & 7.7 & 24.1 & 1.2 & 26.7 & 230.6 & -3.3 &     &     &     \\
CTBP1 & 7.9 & 17.1 & 1.1 & 24.4 & -361.4 & 5.2 &     &     &     \\
PRUNE & 7.6 & 26.2 & 1.1 & 33.6 & 258.6 & -3.7 &     &     &     \\
NELFA & 7.6 & 28.8 & 1.0 & 32.4 & -464.8 & 6.6 &     &     &     \\
MAEA & 7.8 & 14.9 & 1.0 & 33.5 & -418.3 & 6.0 &     &     &     \\
RPS27A & 7.7 & 33.9 & 1.0 & 27.1 & 401.0 & -5.8 &     &     &     \\
TBC1D14 & 7.6 & 21.4 & 0.9 & 34.0 & -510.9 & 7.2 &     &     &     \\
MAN2B2 & 7.7 & 27.7 & 1.1 & 21.9 & -481.9 & 6.8 &     &     &     \\
DGKQ & 7.3 & 31.0 & 1.1 & 36.0 & -552.6 & 7.7 &     &     &     \\
RBFA & 7.6 & -0.7 & 0.9 & 39.7 & -565.2 & 7.8 &     &     &     \\
ACOX3 & 7.7 & 19.0 & 1.2 & 20.3 & -595.1 & 8.2 &     &     &     \\
TCF25 & 7.6 & 27.0 & 1.1 & 33.5 & -578.8 & 8.5 &     &     &     \\
PIGC & 7.6 & 27.7 & 1.1 & 32.6 & 190.4 & -2.7 &     &     &     \\
TOLLIP & 7.2 & 25.6 & 1.0 & 38.7 & -393.2 & 4.9 &     &     &     \\
CREG1 & 7.4 & 28.9 & 1.0 & 44.0 & 198.6 & -2.9 &     &     &     \\
RAB11B & 7.1 & 30.2 & 1.1 & 46.4 & -342.1 & 4.7 &     &     &     \\
CDKN2AIP & 7.8 & 40.2 & 0.9 & 32.3 & -576.4 & 8.5 &     &     &     \\
GAK & 7.9 & 31.3 & 1.1 & 19.4 & -452.0 & 6.5 &     &     &     \\
\hline
	\end{tabular}
\end{sidewaystable}

\begin{sidewaystable}
	\centering
	\caption{Analysis of the lung cancer data using the quantile method: estimates $\times$100.}
	\label{tab6}
	\begin{tabular}{lrrrrr|rrrr}
		\hline
		&\multicolumn{5}{c}{Main effects}&\multicolumn{4}{c}{Interactions}\\
		\hline
		& age & gender & intensity & status & gene & age & gender & intensity & status \\
		\hline
ARHGAP1 & 8.0 & -18.2 & 0.8 & 16.1 & -338.5 & 5.1 &    &    &    \\
B3GALT4 & 7.9 & -8.1 & 1.2 & 14.1 & -58.9 &    &    & 1.5 &    \\
BCL11A & 8.1 & -12.8 & 0.8 & 16.3 & 9.8 &    &    &    &    \\
CDKN1A & 8.1 & -5.3 & 0.7 & 16.9 & -193.6 & 2.5 &    &    &    \\
CHIC2 & 8.3 & -0.3 & 0.9 & -1.0 & -530.8 & 7.8 & 127.8 &    &    \\
CHST10 & 8.1 & -7.3 & 0.8 & 16.6 & 18.9 &    &    &    &    \\
DNM2 & 8.0 & -5.0 & 0.9 & 13.5 & -245.4 & 3.4 &    &    &    \\
FAM65A & 7.8 & 5.0 & 0.8 & 18.5 & -220.7 & 2.7 &    &    &    \\
GATAD1 & 8.1 & -4.2 & 0.9 & 11.8 & -42.0 &    &    &    & 19.6 \\
GGT5 & 8.2 & -14.7 & 0.8 & 14.2 & -152.4 & 1.7 &    &    &    \\
GRB7 & 8.0 & -5.1 & 0.6 & 24.0 & -1.1 &    & 66.4 &    &    \\
HIST1H4D & 8.0 & -13.5 & 0.9 & 16.7 & 13.0 &    &    &    &    \\
LAS1L & 7.8 & -2.1 & 0.8 & 22.9 & -12.8 &    &    &    &    \\
LETM1 & 7.8 & -3.4 & 1.1 & 15.6 & -54.2 &    &    & 1.1 &    \\
MAEA & 7.8 & -3.5 & 1.2 & 13.2 & -36.1 &    &    & 0.6 &    \\
PAPOLG & 8.0 & -16.6 & 0.8 & 12.5 & 16.0 &    &    &    &    \\
PDGFA & 7.9 & 2.5 & 0.8 & 23.1 & -240.6 & 3.5 &    &    &    \\
PDGFRA & 8.0 & 5.7 & 0.6 & 21.1 & -384.8 & 4.9 & 83.9 &    &    \\
PPARD & 8.1 & -5.6 & 0.8 & 13.7 & -270.3 & 3.8 &    &    &    \\
PTAFR & 8.2 & -7.0 & 0.8 & 11.6 & -258.1 & 3.7 &    &    &    \\
RXRB & 7.8 & -5.2 & 1.4 & 7.1 & -71.6 &    &    & 1.3 &    \\
SCARNA9 & 8.1 & -9.5 & 0.7 & 16.7 & 23.7 &    &    &    &    \\
SLC12A4 & 8.0 & -11.2 & 0.8 & 18.3 & -224.0 & 2.7 &    &    &    \\
TMEM204 & 8.1 & -2.0 & 0.8 & 11.5 & -317.9 & 4.2 &    &    &    \\
TOLLIP & 8.0 & -6.3 & 0.9 & 11.7 & -336.9 & 4.8 &    &    &    \\
TOMM5 & 8.0 & -6.9 & 0.8 & 17.8 & 25.2 &    &    &    &    \\
ZNF141 & 7.5 & 24.8 & 1.4 & 8.0 & -105.4 &    & 65.8 & 1.3 &    \\
ZNF761 & 8.0 & -19.2 & 1.0 & 17.0 & -37.4 &    & 83.0 &    &    \\
		\hline
	\end{tabular}
\end{sidewaystable}

\end{spacing}

\end{document}